\documentclass[preprint,showkeys,aip]{revtex4-1}
\usepackage{amsmath}
\usepackage{latexsym}
\usepackage{amssymb}
\usepackage{graphics,epstopdf}
\usepackage{amsmath,amssymb,amsthm}
\usepackage{graphicx}
\usepackage[colorlinks=true, citecolor=blue, urlcolor=blue ]{hyperref}
\usepackage{cleveref}

\begin{document}
	\title{ Information geometry and entanglement under phase-space deformation through nonsymplectic  congruence transformation }
	\author{Shilpa Nandi}
	\email{nandishilpa801325@gmail.com}
	\affiliation{Department of Physics, Shiv Nadar University, Noida, Greater Noida, Uttar Pradesh, India- 201314}
	\author{Pinaki Patra}
	\thanks{Corresponding author}
	\email{monk.ju@gmail.com}
	\affiliation{Department of Physics, Brahmananda Keshab Chandra College, 111/2 B. T. Road, Kolkata, India-700108}

	\date{\today}

	\begin{abstract}
We investigate the relationship between Fisher-Rao (FR) information geometry and Gaussian quantum correlations under nonsymplectic phase-space transformations, with particular emphasis on noncommutative (NC) deformation implemented through the Bopp shift. Using the established affine invariance of the FR-metric, we examine the transformation properties of Gaussian covariance matrices and the positive partial transpose (PPT) criterion under invertible congruence transformations. For a bipartite Gaussian system, we demonstrate that keeping the canonical partial-transpose (PT) matrix fixed in the deformed phase space produces deformation-dependent spectral shifts that may cross the conventional PPT threshold.  When the PT operation is  transported together with the covariance and commutation matrices, the physical PPT spectrum and separability classification remain invariant under local Darboux transformations. We further investigate a two-dimensional anisotropic oscillator in NC phase space, where the deformation parameters modify the effective canonical Hamiltonian and consequently influence the Gaussian ground-state correlations and entanglement between canonical modes. The corresponding FR-geometry reveals that the original  covariance manifold possesses constant negative scalar curvature. In contrast, the pure symmetric normal-form covariance family is intrinsically one-dimensional and flat. These results distinguish the statistical geometry of Gaussian covariance matrices from the physical quantum correlations determined by the Hamiltonian, commutation structure, and subsystem decomposition.
	\end{abstract}
		
	\maketitle
	
	\section{Introduction}
Setting aside debates on the ontology of quantum states, in this paper, we focus on the issue of whether different quantum states remain different after a nonsymplectic transformation under the framework of information geometry \cite{contextuality1,zelinger1,reality1,ontic1,statphysinfo1,complexity1,signal1}. 
Information geometry provides a natural framework for investigating the statistical distinguishability of quantum states \cite{statphysinfo1,complexity1,signal1}. For Gaussian states (GS), the classical Fisher-Rao (FR) geometry associated with their phase-space distributions is completely characterized by the covariance matrix \cite{constrainttheta}. In particular, the FR distance between two Gaussian covariance matrices is invariant under simultaneous invertible congruence transformations. Gaussian entanglement, on the other hand, is characterized not only by the covariance matrix but also by the underlying commutation relations and the tensor-product structure of the system \cite{gaussian1}. This raises an important question: how does the positive partial transpose (PPT) criterion behave under a nonsymplectic congruence transformation that modifies the phase-space commutation structure? We investigate this question by distinguishing two prescriptions for partial transposition: one in which its matrix representation is held fixed and another in which it is consistently transported under the transformation. We demonstrate that these prescriptions can yield different spectral results, whereas the physical PPT classification remains invariant under a consistently transported local congruence transformation.
\\
The geometric formulation of statistical inference, originating in the work of Hotelling and Rao and subsequently developed through the theory of invariant statistical metrics, provides a differential-geometric description of parameterized probability distributions \cite{Hoteling1,Rao1,Chentsov1}. The Fisher information matrix defines a Riemannian metric on the corresponding statistical manifold, allowing the distinguishability of neighboring distributions to be quantified geometrically \cite{Morozova1}. At present, information geometry has become a multidisciplinary branch of statistics having applications in a diverse number of areas, such as statistical physics \cite{statphysinfo1}, complexity theory \cite{complexity1}, signal processing \cite{signal1}, inductive inference \cite{inductive1}, and machine learning \cite{machine1}. In continuous-variable quantum systems, GS offer a particularly tractable setting for this approach because their phase-space distributions are determined by the first moments and covariance matrices. The associated FR geometry can therefore be studied directly in terms of covariance-matrix parameters \cite{infogeo8,infogeo9,optics1,Gaussianref1}.
There has been a considerable amount of attention devoted to understanding their geometric properties \cite{Gaussianref2,Gaussianref3}. The simplicity of GS is that it can be completely determined by its covariance matrix (CVM) $\Sigma$ and first-moment vector \cite{stokes1,stokes2,stokes3}.  First-moment vector can be set to zero for all practical purposes with the help of coordinate shifting. For a family of zero-mean Gaussian phase-space distributions parameterized by $\vartheta$, the classical Fisher information matrix  $g(\vartheta)$ takes the form $g_{\mu\nu}= \frac{1}{2}\mbox{Tr}[\Sigma^{-1}(\partial_\mu \Sigma) \Sigma^{-1}(\partial_\nu \Sigma)]$, where $\partial_\mu $ stands for the partial derivative with respect to the nonzero independent entities $\vartheta_\mu \in \Theta$ of $\Sigma$  \cite{curvature1}. The parameter space $\Theta$ is constrained by the natural constraint imposed by the existence of measurement unsharpness between canonically conjugate observables, through the Robertson-Schr\"{o}dinger uncertainty principle (RSUP) $\Sigma+\frac{i}{2}\Omega\ge 0$, where the  $2n\times 2n$ symplectic matrix $\Omega=[\Omega_{\alpha\beta}]_{\alpha,\beta=1}^{2n}$ encodes the commutation relations $[\hat{X}_\alpha,\hat{X}_\beta]=i\Omega_{\alpha\beta}$  of the phase-space operators $\hat{X}=(\hat{X}_1,...,\hat{X}_{2n})=(\hat{x}_1,...\hat{x}_n,\hat{p}_1,...\hat{p}_n)$ \cite{constrainttheta}. Here we have considered the Planck constant $\hbar=1$, which will be followed throughout the present paper unless otherwise specified. One can endow the parameter space $\Theta$ with a Riemannian metric, namely the FR metric, given by  $G(\vartheta)=\sum_{\mu,\nu}g_{\mu\nu}d\vartheta^\mu \otimes d \vartheta^\nu$, which induced a Riemannian manifold $\mathcal{M}=(\Theta, G(\vartheta))$ \cite{infogeo1}. The FR distance between two GS, corresponding to CVM $\Sigma_1$ and $\Sigma_2$ is defined as $d_{FR}(\Sigma_1,\Sigma_2)=\sqrt{\frac{1}{2}\sum_{j=1}^{2n}\log^2\lambda_j(\Sigma_1^{-1}\Sigma_2)}$, where $\lambda_j(\Sigma_1^{-1}\Sigma_2)$ are the $j^{th}$ largest generalized eigenvalues, which are solutions of $\mbox{Det}(\Sigma_1-\lambda\Sigma_2)=0$ \cite{metricFR1}.  $d_{FR}$ is invariant under any congruent transformation $\tilde{\Sigma}=\Upsilon\Sigma \Upsilon^T$, where $\Upsilon\in GL(2n,\mathbb{R})$ is an invertible $2n\times 2n$ matrix in $\mathbb{R}$ \cite{metricFR2,metricFR3}. \\
Noncommutative (NC) phase-space models provide a useful theoretical setting for examining modifications of the canonical commutation relations. In such models, position and momentum operators may satisfy nonvanishing commutators between different spatial components, characterized by deformation parameters $\theta$ and $\eta$ \cite{ncs1,ncs2,nandi1,Moyal1,Moyal2}. The Bopp transformation relates these NC variables to canonical phase-space coordinates through an invertible, generally nonsymplectic linear transformation. This construction provides a tractable framework for investigating how covariance matrices, commutation structures, and Gaussian PPT criteria transform under phase-space deformation. Related noncommutative structures also arise in effective descriptions of physical systems, including the guiding-center coordinates of charged particles in magnetic fields \cite{Hall1,Hall2,ncs3,ncssim1,Gouba1,NCent1,NCent2,NCent3,NCent4}.\\
The central objective of the present work is to clarify the relationship between FR isometry and Gaussian PPT classification under nonsymplectic phase-space congruence transformations. We first consider a bipartite Gaussian state and apply a local Bopp transformation to its covariance matrix and commutation structure. When the matrix representing partial transposition (PT) is held fixed, the resulting spectral quantities acquire an explicit dependence on the NC deformation parameters, and the corresponding minimum eigenvalue may cross the conventional PPT threshold. However, when PT is consistently transported with the phase-space transformation, the PPT condition remains invariant \cite{ppt1}. This establishes that the fixed-transpose spectral deformation must be distinguished from a change in physical entanglement. We subsequently examine a two-dimensional anisotropic oscillator whose Hamiltonian depends on the NC parameters and investigate the resulting Gaussian correlations and information geometry in canonical coordinates.   
\\
The paper is organized as follows. In Section-II, we review the Fisher-Rao geometry of Gaussian covariance matrices and the uncertainty and separability conditions. Section-III establishes the invariance of the Fisher-Rao distance under invertible congruence transformations. In Section-IV, we investigate a bipartite Gaussian system under noncommutative phase-space deformation, comparing fixed and transported prescriptions for partial transposition and their corresponding spectral properties. Section-V considers a two-dimensional anisotropic oscillator in noncommutative phase space, examining its Gaussian ground state, covariance matrix, separability properties relative to canonical modes, and associated information geometry. Finally, we summarize the principal results and their implications.
\section{Geometry induced by Gaussian states}
A $n$-mode quantum system is described by its density opeartor $\hat{\rho}(\hat{\xi})$ on the Hilbert space $\mathcal{H}=L^2(\mathbb{R})^{\otimes n}$. Here $\hat{\xi}=(\hat{\xi}_1,...,\hat{\xi}_{2n})=(\hat{q}_1,...,\hat{q}_n,\hat{p}_1,...,\hat{p}_n)$ is the phase-space co-ordiante. The canonical positions ($\hat{q}_k$) and momenta ($\hat{p}_k$) satisfy the canonical commutation relations
\begin{eqnarray}
	\left[\hat{\xi}_k,\hat{\xi}_l\right]=i \Omega_{kl},\; k,l=1,..,2n,
\end{eqnarray} 
where $\Omega_{kl}$ is the $kl^{th}$ element of the symplectic matrix
\begin{eqnarray}
	\Omega= \left(\begin{array}{cc}
		0 & \mathbb{I}_n\\
		-\mathbb{I}_n & 0
	\end{array}\right),
\end{eqnarray}
$\mathbb{I}_n$ being the $n\times n$ identity matrix. Phase-space representation of $\hat{\rho}$ is given by the quasiprobability distribution, namely the Wigner distribution, defined as
\begin{equation}\label{Wignerdefn}
	W(\xi)=(1/\pi)^N \int e^{2i p_ky_k} \rho (q_1+y_1,q_1-y_1,...., q_N+y_N,q_N-y_N) dy_1...dy_N,
\end{equation}
which can be considered as originating from the characteristic function $\chi_{\hat{\rho}}(\tau)$ through the Fourier transformation
\begin{eqnarray}
	W(\xi)=\int e^{-i\xi^T\tau}\chi_{\hat{\rho}}(\tau)d\tau,
\end{eqnarray}
where $\chi_{\hat{\rho}}(\tau)$ is the expectation value of the displacement operator $\hat{D}(\xi)$. In particular,
\begin{equation}
	\chi_{\hat{\rho}}(\tau)=\mbox{Tr}(\hat{\rho }\hat{D}(\xi)),\; \mbox{with}\; \hat{D}(\xi)= e^{i(q_k\hat{q}_k+p_k\hat{p}_k)}.
\end{equation}
Here $X^T$ stands for matrix transposition of $X$.  We have used Einstein's summation convention (sum over repeated indices, e.g., $p_ky_k=\sum_{j=1}^{N}p_jy_j$), which will be followed throughout the paper unless otherwise specified. Gaussian functions are one of the eigenvectors of Fourier transformations. We define, Gaussian states are those for which the characteristic function is a Gaussian function of the phase space coordinates. In particular, for a Gaussian state $\hat{\rho}$, the characteristic function reads
\begin{equation}
	\chi_{\hat{\rho}}(\xi)=e^{-1/2\xi^T\Sigma \xi +i\mu^T\xi},
\end{equation}
where $\mu=(\mu_1,...,\mu_{2N})$ is the first moment vector and $\Sigma=[\Sigma_{ij}]_{i,j=1}^{2N}$ is the covariance matrix, which are defined by 
\begin{equation}
	\mu_j=\langle \xi_j\rangle, \;
	\Sigma_{kl}=\frac{1}{2}\langle \left\{\xi_k,\xi_l\right\}\rangle -  \langle \xi_k\rangle \langle \xi_l\rangle,
\end{equation}
where the expectation values ($	\langle \mathcal{O}\rangle$) of operators ($\mathcal{O}$) are taken over the phase-space distribution $W(\xi)$, and defined as 
\begin{equation}
	\langle \mathcal{O}\rangle =	\langle \mathcal{O}\rangle_W=\int \mathcal{O}W(\xi) d^{2N}\xi.
\end{equation}
Note that, the expectation values over the state $\hat{\rho}$ are the same as the expectation values over Wigner distributions, i.e., $\langle \mathcal{O}\rangle_W=\langle \mathcal{O}\rangle_{\hat{\rho}}$. Now, the corresponding Gaussian state can be written as
\begin{equation}
	W= \frac{e^{-\frac{1}{2}(\xi-\mu)^T \Sigma^{-1}(\xi-\mu)}}{(2\pi)^N \sqrt{\Delta_{\Sigma}}},
	\;\mbox{with}\; \Delta_\Sigma=\mbox{Det}(\Sigma).
\end{equation}
Without loss of generality, one chooses the first-moment vector $\mu=0$ throughout the present paper. Now, we parametrize the associated Gaussian probability density function $P(\xi)$,  by the nonzero entities of the covariance matrix as follows.
\begin{equation}
	\mathcal{S}:=\left\{ P(\xi)\equiv P(\xi;\vartheta)=\frac{e^{-\frac{1}{2}\xi^T \Sigma^{-1}(\vartheta)\xi}}{(2\pi)^N\sqrt{\mbox{det} \Sigma(\vartheta)}},\vert \vartheta \in \Theta \subseteq \mathbb{R}^m \right\},
\end{equation}
where the parametrization $\vartheta=(\vartheta_1,....,\vartheta_m)$ is provided through elements  of $\Sigma$ as
\begin{equation}
	\vartheta_l=\Sigma_{\mu\nu},\; \mbox{with}\; l=\sum_{r=0}^{\mu-2} (2n-r)+\nu-\mu+1,\; 1\le l\le m.
\end{equation}
We note that $\vartheta \mapsto P(.;\vartheta)$ is injective and $\Theta \subseteq \mathbb{R}^m$ is obtained from specific constraint on $\Sigma(\vartheta)$. For a classical distribution, a bonafide covariance matrix is positive definite. Due to the existence of natural measurement unsharpness between the canonically conjugate observables (such as $\hat{q_k}$ and $\hat{p}_k$) in quantum mechanics,  a covariance matrix corresponding to a quantum state satisfies Robertson Schr\"{o}dinger uncertainty principle (RSUP)
\begin{equation}\label{RSUP}
	\Sigma + \frac{i}{2}\Omega\ge 0.
\end{equation}
Here the notation $A\ge 0$ stands for the matrix $A$ is positive semidefinite. We write $A>B$ if $A-B$ is positive definite.
Thus, the parameter space $\Theta$ for quantum states  is given by
\begin{equation}\label{thetaquantum}
	\Theta_{\mbox{quantum}}:= \left\{\vartheta\in \mathbb{R}^m\vert \Sigma(\vartheta)+\frac{i}{2}\Omega\ge 0\right\}.
\end{equation}
From the definition, it follows that the  transposition operation ($\Lambda$) corresponds to the transformation of the sign to all momentum ($p_k$)  and leaves the coordinates ($q_k$) unchanged
\begin{equation}
	(q_1,...,q_n,p_1,...,p_n)\mapsto \Lambda(q_1,...,q_n,p_1,...,p_n)=(q_1,...,q_n,-p_1,...,-p_n).
\end{equation}
For a composite Gaussian system with two subsystems $A$ and $B$, the positive partial transposition (PPT) operation $\Lambda_A$ in the system $A$  is defined by $\Lambda_A=\Lambda\oplus \mathbb{I}_B$, and similarly $\Lambda_B=\mathbb{I}_A\oplus\Lambda$, where $\mathbb{I}_{A(B)}$ corresponds to the identity operator of the subsystem $A(B)$. In other words, PPT operation performs transposition only on the corresponding subsystem, leaving other subsystems unaffected. If $\Sigma_A$ and $\Sigma_B$ are the covariance matrices corresponding to the subsystems $A$ and $B$, and $\Sigma$ is the covariance matrix of the whole composite system, then  a Gaussian state described by the covariance matrix $\Sigma$ is separable if and only
if there exist covariance matrices $\Sigma_A$ and $\Sigma_B$ such that $\Sigma \ge \Sigma_A\oplus \Sigma_B$. For two-mode Gaussian systems, this criteria simplifies to the following. 
A $4\times 4$ covariance matrix $\Sigma$ satisfying ~\eqref{RSUP}, describes a separable state if and only if
\begin{equation}\label{PPTforB}
	\Sigma'+\frac{i}{2}\Omega\ge 0, \; \mbox{where}\; \Sigma'=\Lambda_B \Sigma \Lambda_B,\; \mbox{with}\; 
	\Lambda_B(q_1,q_2,p_1,p_2)=(q_1,q_2,p_1,-p_2),
\end{equation}
or equivalently for the case of PPT with respect to $A$. ~\eqref{PPTforB} provides the following constraint for the set of separable states of bipartite Gaussian state
\begin{equation}\label{thetaseparable}
	\Theta_{\mbox{separable}}:= \left\{\vartheta\in \mathbb{R}^m\vert \Sigma(\vartheta) \ge \Sigma_A\oplus \Sigma_B\right\}.
\end{equation}
Therefore, the parameter space for the set of entangled states is 
\begin{equation}\label{thetaentangled}
	\Theta_{\mbox{entangled}}:= \Theta_{\mbox{quantum}}-\Theta_{\mbox{separable}}.
\end{equation}  
If we restrict our discussion for two-mode covariance matrix $\Sigma(\vartheta)$, we note that the standard-form covariance matrix parametrization of $\Sigma(\vartheta)$ is realized through its canonical form 
\begin{eqnarray}\label{Sigma}
	\Sigma=\left(\begin{array}{cccc}
a & 0 & c & 0\\
0 & a & 0 & d \\
c & 0 & b & 0\\
0 & d & 0 & b
	\end{array}\right).
\end{eqnarray}
In that case, the state is parametrized by four parameters $\vartheta=(a,b,c,d)\in \mathbb{R}^4$. If the state has EPR-type correlation structure then $d=-c$, it is symmetric  for $a=b$. The one-parameter pure squeezed-state family is obtained for the additional constraint $a^2-c^2=1/4$. \\
For  a bonafide covariance matrix ~\eqref{Sigma}, the explicit form of the constraint~\eqref{thetaquantum} reads
\begin{eqnarray}
\Theta_{\mbox{quantum}} = \{   a>1/2, \;1/2<b<a, \; c^2 <\delta_1/4,\; \delta_{-}\le d\le \delta_{+} 
  \} \nonumber \\
  \cup
\{  a>1/2,\; 1/2<a<b, \;
  c^2 < \delta_2/4, \;\delta_{-}\le d\le \delta_{+} \},
\end{eqnarray}
where we have used following notations for parameters.
\begin{eqnarray}
	\delta_1=\delta_0-\nu_0,\; \delta_2=\delta_0-\nu_0^{-1},\; \delta_3= (\sqrt{\delta_0}+1/\sqrt{\delta_0})^2 - (\sqrt{\nu_0}+1/\sqrt{\nu_0})^2, \\
	\Delta_0 = c^2 +\frac{1}{4}\delta_0 (\delta_0 -4c^2)(\delta_3-4c^2), \; \delta_{\pm}=\frac{-c\pm \sqrt{\Delta_0}}{\delta_0-4c^2},\;
	\mbox{with}\; \nu_0=a/b,\; \delta_0=4ab.
\end{eqnarray}
For a separable state the PPT criterion ~\eqref{PPTforB} or equivalently $\Sigma+\frac{i}{2}\Omega'\ge0$, with $\Omega'=\Lambda_B\Omega\Lambda_B$, reduces the constraint ~\eqref{thetaseparable} as follows.
\begin{eqnarray}
	\Theta_{\mbox{separable}} = \{   a>1/2, \; b>1/2, \; -\sqrt{\delta_3}<2c<0,\; \delta_{-}\le d\le -\delta_{-} 
	\} \nonumber \\
	\cup
	\{  a>1/2,\; b>1/2, \; 0<2c < \sqrt{\delta_3}, \;- \delta_{+}\le d\le \delta_{+} \}.
\end{eqnarray}
RSUP ~\eqref{RSUP} can be equivalently stated in terms of the symplectic spectrum of $\Sigma$, i.e., the ordinary eigenvalues of $2i\Omega^{-1}\Sigma$. All the eigenvalues of $2i\Omega^{-1}\Sigma$ are greater than one, and they are  Williamson invariant, i.e., they remain invariant under symplectic transformation. By Williamson's theorem, one can show that the RSUP is equivalent to the statement $\nu_{-}\ge1$, where $\nu_{-}$ is the smallest eigenvalue of   $2i\Omega^{-1}\Sigma$. \\
The Fisher information matrix $g(\vartheta)=[g_{\mu\nu}]_{\mu,\nu=1}^{m}$ for the family of Gaussian states $\mathcal{S}$ at a point $\vartheta\in \Theta$ is defined by
\begin{equation}\label{gmunudefn}
	g_{\mu\nu}(\vartheta)= \int_{\mathbb{R}^{2N}} P(\xi;\vartheta)\partial_\mu \ln P(\xi;\vartheta)\partial_\nu \ln P(\xi;\vartheta)dx.
\end{equation}
Here the notation $\partial_\mu$ stands for partial derivative with respect to $\vartheta_\mu$. 
In terms of the covariance matrix, ~\eqref{gmunudefn} can be rewritten as
\begin{equation}\label{gmunuV}
	g_{\mu\nu}= \frac{1}{2}\mbox{Tr}[\Sigma^{-1}(\partial_\mu \Sigma)\Sigma^{-1}(\partial_\nu \Sigma)],
\end{equation}
which is computationally simpler than that of ~\eqref{gmunudefn}.
The positive semidefinite matrix $g(\vartheta)$ endows the parameter space $\Theta$ with a Riemannian metric (the Fisher-Rao (FR) metric) 
\begin{equation}
	G(\vartheta):= \sum_{\mu\nu} g_{\mu\nu}(\vartheta)d\vartheta^{\mu}\otimes d\vartheta^{\nu}.
\end{equation}
The FR-metric provides a natural Riemannian structure on the parameter space of Gaussian covariance matrices. Its geometric properties characterize the statistical distinguishability of neighboring Gaussian states and can be investigated through the corresponding metric tensor, connection coefficients, curvature, and geodesic distance. In the present work, our primary interest is the behavior of this information-geometric structure under invertible congruence transformations and its relationship with the positive partial transpose criterion for Gaussian states.\\
With the parametrization $\theta=(\vartheta_1,\vartheta_2,\vartheta_3,\vartheta_4)=(a,b,c,d)$, using ~\eqref{gmunuV}, we get
following components of FR-metric for the covariance matrix ~\eqref{Sigma}.
\begin{eqnarray}
	g_{11} &=& \frac{b^2}{\Delta_\Sigma}\left[ 1 + \frac{(c^2-d^2)^2}{2\Delta_\Sigma}\right],\; 
	g_{12}= \frac{1}{2}\left(\frac{c^2}{\Delta_c^2}+\frac{d^2}{\Delta_d^2}\right),
	g_{13}=-\frac{bc}{\Delta_c^2},\nonumber \\
	 g_{14}&=& -\frac{bd}{\Delta_d^2},\; 
	g_{22}= \frac{a^2}{\Delta_\Sigma}\left[1+ \frac{(c^2-d^2)^2}{2\Delta_\Sigma}\right],\;
	g_{23}=-\frac{ac}{\Delta_c^2},\nonumber \\
	g_{24} &=& -\frac{ad}{\Delta_d^2},\;
	g_{33}=\frac{ab+c^2}{\Delta_c^2},\; g_{34}=0,\; g_{44}=\frac{ab+d^2}{\Delta_d^2},
\end{eqnarray}
with
\begin{equation}
	\Delta_c=ab-c^2,\; \Delta_d=ab-d^2,\; \Delta_\Sigma=\mbox{Det}(\Sigma)=\Delta_c\Delta_d.
\end{equation}
The determinant of the FR-information matrix $g=[g_{ij}]$ is given by
\begin{equation}\label{Detg}
	\Delta_g=\mbox{Det}(g)=\frac{1}{4\Delta_\Sigma^3}[4a^2b^2-(c^2+d^2)^2].
\end{equation}
Note that, for the EPR-type correlation ($d=-c$), the FR-metric is reduced to the following three parameter family
\begin{equation}
	g^{(3)}_{EPR}=\frac{1}{(ab-c^2)^2}\left( \begin{array}{ccc}
b^2 & c^2 & -2bc\\
c^2 & a^2 & -2ac \\
-2bc & -2ac & 2(ab+c^2)
	\end{array}\right),
\end{equation}
whereas for the EPR-type symmetric state family ($a=b,\;d=-c$), the two-parameter family metric reads
\begin{equation}
g^{(2)}_{sym}= \frac{2}{(a^2-c^2)^2}\left( \begin{array}{cc}
a^2+c^2 & -2ac \\
-2ac & a^2+c^2
\end{array}\right).
\end{equation}
Note that for pure state we have purity condition $a^2-c^2=1/4$, which provides a convenient parametrization through two-mode squeezing parameter (squeezing amplitude) $r$ to express the FR-metric in a one-parameter family as follows.
\begin{equation}
	a=\frac{1}{2}\cosh 2r ,\ c=\frac{1}{2}\sinh 2r \implies ds_{FR}^2=8dr^2.
\end{equation}
In other words, the pure state submanifold is intrinsically flat (Ricci-scalar $R=0$).
Thus we see that, the information-geometric sensitivity to a change in the squeezing parameter ($r$) is constant along this one-dimensional family. An infinitesimal change $dr$ produces $ds=2\sqrt{2}dr$. Consequently the distance between two pure states (characterized by $r_1$ and $r_2$) along this one dimensional manifold is given by $d_{FR}^{pure}(r_1,r_2)= 2\sqrt{2}\vert r_2-r_1\vert$.
\section{Distance between general bipartite gaussian states}
The notion of distance between two covariance matrices with the FR-metric by
\begin{equation}
	d_{FR}(\Sigma_1,\Sigma_2)= \sqrt{\frac{1}{2}\sum_{j=1}^{2N}\log^2 \lambda_j(\Sigma_1^{-1}\Sigma_2)},
\end{equation} 
where the notation $\lambda_j(M)$ stands for the $j^{th}$ generalized eigenvalue of $M$, i.e., $\lambda_j$'s are the solutions of the equation  $\mbox{Det}(\Sigma_2-\lambda\Sigma_1)=0$.
$\Sigma_1^{-1}\Sigma_2$ is not symmetric in general. However, $\Sigma_1^{-1/2}\Sigma_2 \Sigma_1^{-1/2}$ is symmetric and positive definite, and $\lambda_j(\Sigma_1^{-1}\Sigma_2)=\lambda_j(\Sigma_1^{-1/2}\Sigma_2 \Sigma_1^{-1/2})$.
Therefore, one can equivalently define FR-metric as
\begin{equation}
	d_{FR}(\Sigma_1,\Sigma_2)= \sqrt{\frac{1}{2}\sum_{j=1}^{2N}\log^2 \lambda_j(\Sigma_1^{-1/2}\Sigma_2 \Sigma_1^{-1/2})}.
\end{equation} 
For instance, for he covariance matrix of the form ~\eqref{Sigma}, i.e., 
$\Sigma_j=\left(\begin{array}{cc}
	a_j\mathbb{I}_2 & \mbox{diag}(c_j,d_j)\\
	\mbox{diag}(c_j,d_j) & b_j\mathbb{I}_2
\end{array}\right)$, $j=1,2$, the explicit form of eigenvalues $spec(\Sigma_1^{-1}\Sigma_2)=\{\lambda_{c,\pm},\lambda_{d,\pm}\}$ reads
\begin{equation}
	\lambda_{c,\pm}=\frac{1}{2\Delta_{c,1}}(K_c\pm \sqrt{K_c^2-4\Delta_{c,1}\Delta_{c,2}}),\;
	\lambda_{d,\pm}=\frac{1}{2\Delta_{d,1}}(K_d\pm \sqrt{K_d^2-4\Delta_{d,1}\Delta_{d,2}}),
\end{equation}
where
\begin{eqnarray}
	\Delta_{c,i}=a_ib_i-c_i^2,\; \Delta_{d,i}=a_ib_i-d_i^2,\;i=1,2.\\
	K_c=a_1b_2+a_2b_1-2c_1c_2,\; K_d=a_1b_2+a_2b_1-2d_1d_2.
\end{eqnarray}
Consequently, the distance function reads
\begin{equation}
	d(\Sigma_1,\Sigma_2)=\sqrt{\frac{1}{2} (\ln^2\lambda_{c,+} + \ln^2\lambda_{c,-} + \ln^2\lambda_{d,+} + \ln^2\lambda_{d,-})}.
\end{equation}
One can note that, if we change index $1\leftrightarrow 2$, the eigenvalues changes as $\lambda_j\to \lambda_j^{-1}$. Moreover, if $\Sigma_1=\Sigma_2$, then $K_c=2\Delta_c,\; K_d=2\Delta_d$, which gives $\lambda_{c,\pm}=\lambda_{d\pm}=1$. In other words, the consistency conditions for a metric holds, as
\begin{equation}
	d_{FR}(\Sigma_1,\Sigma_2)=d_{FR}(\Sigma_2,\Sigma_1),\; d_{FR}(\Sigma,\Sigma)=0.
\end{equation}
An important property of this distance is its invariance under a simultaneous invertible congruence transformation. Let us consider that the congruence transformation $\Upsilon\in GL(2n,\mathbb{R})$ transforms two covariance matrix as
\begin{equation}
	\Sigma_1'=\Upsilon\Sigma_1\Upsilon^T,\; \Sigma_2'=\Upsilon\Sigma_2\Upsilon^T;\;\; \mbox{Det}(\Upsilon)\neq 0.
\end{equation}
The generalized eigenvalues for the transformed pair are obtained from
\begin{equation}
	\mbox{Det}(\Sigma_2'-\lambda\Sigma_1')= \mbox{Det}[\Upsilon(\Sigma_2-\lambda\Sigma_1)\Upsilon^T]=(\mbox{Det}(\Upsilon))^2\mbox{Det}(\Sigma_2-\lambda\Sigma_1)=0,
\end{equation}
which means the set of eigenvalues remain unchanged after any invertible congruence $GL(2n,\mathbb{R})$. In other words, every invertible congruence transformation acts as a isometry of the Fisher-Rao geometry of the manifold of positive-definite covariance matrices.\\
It is important to distinguish the statistical geometrical statement of invariance of FR-metric, from the general separability criteria of Gaussian states. In particular, FR-metric solely depends on covariance matrix, whereas the separability criteria depends additionally on the symplectic structure. Thus the relevant structure for separability  condition is
\begin{equation}
	(\Sigma,\Omega) \to (\Upsilon\Sigma \Upsilon^T, \Upsilon\Omega \Upsilon^T).
\end{equation}
Consequently, invariance of FR-distance under congruence does not, by itself imply invariance of Gaussian PPT/separability classification when underlying canonical structure is deformed. Additional constraint on the partial transposition (PT) operation is indeed provide different results. In particular, in the subsequent section, we  demostrate that the transformation rule of PT operation is crucial for PPT-separability of Gaussian states.
\section{Bipartite Gaussian states under noncommutative deformation of phase-space }  
Suppose, we have a bipartite system $\mathcal{H}=\mathcal{H}_A\otimes\mathcal{H}_B$ consisting of $A$ and $B$ in two dimensional background space (i.e., each subsystem has four dimensional phase space). The composite system  has eight-dimensional phase space with co-ordinates $\hat{\xi}=(\hat{x}_{1}^{A}, \hat{x}_{2}^{A},\hat{p}_{1}^{A},\hat{p}_{2}^{A},\hat{x}_{1}^{B}, \hat{x}_{2}^{B},\hat{p}_{1}^{B},\hat{p}_{2}^{B})^T$ having the canonical commutation relations encoded in the symplectic matrix
\begin{eqnarray}
	\Omega=\Omega_A\oplus\Omega_B=\mbox{Diag}(\Omega_A,\Omega_B),\; \mbox{with}\; 
	\Omega_A=\Omega_B=J_4=\left(\begin{array}{cc}
0 &\mathbb{I}_2 \\
-\mathbb{I}_2 & 0
	\end{array}\right).
\end{eqnarray} 
If, $A$ and $B$ share a composite state
\begin{equation}
	F(\xi)=\frac{1}{(2\pi)^4\sqrt{\mbox{Det}(\Sigma)}}e^{-\frac{1}{2}\xi^T\Sigma^{-1}\xi}, 
\end{equation}
then with the help of local symplectic transformation ($\mathcal{S}_{AB}=\mathcal{S}_A\oplus \mathcal{S}_B$), one can express the covariance matrix $\Sigma_{AB}$ in the  canonical form
\begin{eqnarray}\label{Sigmacanonical}
	\Sigma_{AB}=\left( \begin{array}{cc}
\sigma_{AA} & \sigma_{AB}\\
\sigma_{AB}^T & \sigma_{BB}
	\end{array}\right),
\end{eqnarray}
with
\begin{eqnarray}
	\sigma_{kk}=\bigoplus_{j=1}^{2}\sigma_{kj}\mathbb{I}_2,\; k=A,B.\\
	\sigma_{AB}=\left(\begin{array}{cc}
\mbox{Diag}(\sigma_{11},\sigma_{22}) & \mbox{Diag}(\sigma_{13},\sigma_{24})\\
\mbox{Diag}(\sigma_{13},\sigma_{24}) & \mbox{Diag}(\sigma_{33},\sigma_{44}
	\end{array}\right).
	\end{eqnarray}
Thus the composite bipartite system in  two dimensional background space can be expressed through the ten parameters 
$ (\sigma_{A1},\sigma_{A2},\sigma_{B1},\sigma_{B2},\sigma_{11},\sigma_{22},\sigma_{13},\sigma_{24},\sigma_{33},\sigma_{44}).
$
Since our aim is to focus on entanglement induced by the congruence, we restrict to a tractable toy model, provided by Bastos et al. \cite{NCent1} of a symmetric family of Gaussian states parametrized by $m,n,b $ as follows. 
\begin{eqnarray}
\sigma_{A1}=\sigma_{A2}=\sigma_{B1}=\sigma_{B2}=b/2,\;\mbox{with}\; b>0.\\
\sigma_{11}=\sigma_{22}=-\sigma_{33}=-\sigma_{44}=nb/2,\\
\sigma_{13}=-\sigma_{24}=mb/2,\; \mbox{with}\; m,n\in \mathbb{R}.
\end{eqnarray}
Note that, the states considered here are generally mixed.
The covariance matrix ~\eqref{Sigmacanonical} is rewritten in terms of the parameters $m,n,b$ as
\begin{eqnarray}
	\Sigma_{AB}=\frac{b}{2}\left(\begin{array}{cc}
\mathbb{I}_4 & \gamma^T\\
\gamma & \mathbb{I}_4
	\end{array}\right),\;
\mbox{with}\;
\gamma=\left(\begin{array}{cc}
n\mathbb{I}_2 & m\sigma_z \\
m\sigma_z & -n\mathbb{I}_2
\end{array}\right).
\end{eqnarray}
Here, we have used the following reprentation for the Pauli matrices.
\begin{eqnarray}
	\sigma_x=\left(\begin{array}{cc}
0 & 1\\
1& 0
	\end{array}\right),\; 
	\sigma_y=\left(\begin{array}{cc}
	0 & -i\\
	i& 0
\end{array}\right),\;
	\sigma_z=\left(\begin{array}{cc}
	1 & 0\\
	0& -1
\end{array}\right).
\end{eqnarray}
For simplicity let us  choose 
\begin{equation}
b=(1+R)/(1-R),\; \mbox{with}\;	R=\sqrt{m^2+n^2}.
\end{equation}
Since $b>0$, we have $R<1$. In other words, $m,n$ lies inside a unit circle, centered at the origin. For $b>0$ and $R<1$, the covariance matrix $\Sigma_{AB}$ is positive definite.
\subsection{Information geometry}
The parameter space of states is two-dimensional, characterized by $m,n$.  If we reparametrize $m,n$ as
\begin{equation}
	m=R\cos\phi,\; n=R\sin\phi,
\end{equation}
direct computation yields the FR-metric as
\begin{eqnarray}
	g(R,\phi)=\mbox{Diag}(g_{RR},g_{\phi\phi}),\;\; \mbox{with}\\
	 g_{RR}=\frac{4(R^2-4R+5)}{(1-R^2)^2},\;
	g_{\phi\phi}=\frac{4R^2}{1-R^2}.
\end{eqnarray}
Note that the metric only depends on $R$, and not on the angular variable $\phi$. Moreover,
both $\lim_{R\to 1-} g_{RR} \to \infty$ and $\lim_{R\to 1-} g_{\phi\phi} \to \infty$. The boundary $R=1$ becomes strongly stretched in information-geometric co-ordinates.
From the metric  $ds_{FR}^2=g_{RR}dR^2 + g_{\phi\phi}d\phi^2$, one can compute Christoffel symbols $\Gamma_{\mu\nu}^{\alpha}$ and Ricci tensors $R_{\mu\nu}$ ($\mu,\nu,\alpha=R,\phi$), which reads
\begin{eqnarray}
	\Gamma_{RR}^R &=& \frac{R^3-6R^2+11R-2}{(1-R^2)(R^2-4R+5)},\\
		\Gamma_{R\phi}^\phi &=&\Gamma_{\phi R}^\phi =\frac{1}{R(1-R^2)},\;\; \Gamma_{\phi\phi}^R=-\frac{R}{R^2-4R+5}.
\\
	R_{RR} &=& -\frac{2(R^3-3R^2+2R+1)}{R(1-R^2)^2(R^2-4R+5)},\;
	R_{\phi\phi}=-\frac{2R(R^3-3R^2+2R+1)}{(1-R^2) (R^2-4R+5)^2}.
\end{eqnarray}
All other components are zero. The scalar curvature $\mathcal{R}=g^{\mu\nu}R_{\mu\nu}$ is given by
\begin{equation}
	\mathcal{R}(R)=-\frac{R^3-3R^2+2R+1}{R(R^2-4R+5)^2}<0;\;\; 0<R<1.
\end{equation}
It is worth noting that the divergence $\lim_{R\to 0}\mathcal{R}\to -\infty$ is not a physical singularity or entanglement effect. Rather, it is the manifestation of the nonsmooth constraint $b=(1+R)/(1-R)$.
\\
We now investigate the transformation properties of Gaussian PPT criteria under local nonsymplectic congruence transformations, distinguishing the fixed PT spectral prescription from the consistently transported physical PT.
\subsection{Covariance of Gaussian PPT under local congruence transformation}
If $\pm  \lambda_{k}\in\mathbb{R},\; (k=1,...4)$ are the eigenvalues of $2i\Omega^{-1}\Sigma_{AB}$, then the diagonal entities (symplectic eigenvalues of $\Sigma_{AB}$) $\mbox{Diag}(\sigma_z,\sigma_z,\sigma_z,\sigma_z)\mbox{Diag}(\lambda_1,-\lambda_1,\lambda_2,-\lambda_2,\lambda_3,-\lambda_3,\lambda_4,-\lambda_4)$ are the Williamson invariants. For a covariance matrix, the smallest Williamson invariant has to be greater than one, which is equivalent statement to RSUP $\Sigma_{AB}+(i/2)\Omega\ge 0$. In the present case, all the symplectic eigenvalues are same and equal to
\begin{equation}\label{RSUPspec}
	\lambda_{\Sigma,min}=b\sqrt{1-R^2}=\frac{(1+R)^{3/2}}{\sqrt{1-R}}>1,\;\forall R\in (0,1).
\end{equation}
Therefore $\Sigma_{AB}$ is a valid covariance matrix for a quantum system.  Now, in order to determine the separability of the state let us perform the partial transposition (PT) operation on the party $B$. The PT transformation changes the sign of momentum co-ordinates of $B$:
\begin{equation}\label{PToperation}
	\xi\mapsto \xi'=\Lambda \xi,\; \mbox{with}\; \Lambda=\Lambda_A\oplus\Lambda_B,\; \Lambda_A=\mathbb{I}_4,\; \Lambda_B=\mbox{Diag}(\mathbb{I}_2,-\mathbb{I}_2).
\end{equation} 
Under PT, the covariance matrix of a separable state will be transformed into a bonafide covariance matrix. In other words,
\begin{equation}\label{PTcan}
	\Sigma'_{AB}+(i/2)\Omega\ge 0, \; \mbox{where}\; \Sigma'_{AB}=\Lambda\Sigma_{AB}\Lambda^T.
\end{equation}
The symplectic invariants of $\Sigma'_{AB}$ are 
\begin{equation}\label{PTspec}
	\lambda_{\Sigma'}=b(1\pm R).
\end{equation}
The smallest symplectic invariant ($	\lambda_{\Sigma',min}=b(1-R)= 1+R>1$) is always greater than one. Since PPT criteria is not sufficient for eight dimensional phase-space, we shall call the initial state as a PPT-state, which satisfies PPT criteria. We wish to explore how a congruence transformation affects PPT criteria. For a physically motivated example, let us consider the congruence transformation
\begin{eqnarray}
	\hat{\tilde{\xi}}=\Upsilon\hat{\xi},\; \mbox{with}\;	\Upsilon=\mbox{Diag}(\Upsilon_A,\Upsilon_B),\;  \Upsilon_A=\Upsilon_B= \left( \begin{array}{cc}
		\mathbb{I}_2 & -i\frac{\theta}{2}\sigma_y \\
		i\frac{\eta}{2}\sigma_y & \mathbb{I}_2
	\end{array}\right).
\end{eqnarray}
Since $\Upsilon$ is invertible we have the following constraints on the parameters
\begin{equation}
	\Delta_\Upsilon=(1-\theta\eta/4)^4\neq 0,\; \Delta_{\Upsilon_A}=\Delta_{\Upsilon_B}=(1-\theta\eta/4)^2 \neq 0. 	
\end{equation}
Here we have used the notation $\Delta_U=\mbox{Det}(U)$, and we consider the Planck's constant $\hbar=1$.
The canonical commutation relations are encoded in the deformed symplectic matrix
\begin{eqnarray}
	\tilde{\Omega}=\Upsilon\Omega \Upsilon^T=\mbox{Diag}(\tilde{\Omega}_A,\tilde{\Omega}_B),
\end{eqnarray}
where
\begin{eqnarray}
	\tilde{\Omega}_A=\tilde{\Omega}_B= \Upsilon_A \Omega_A \Upsilon_A^T=\left(\begin{array}{cc}
		i\theta \sigma_y & \hbar_e\mathbb{I}_2\\
		-\hbar_e\mathbb{I}_2 & i\eta \sigma_y
	\end{array}\right),\; \mbox{with}\; \hbar_e=1+\theta\eta/4.
\end{eqnarray}
which corresponds to the noncommutative (NC) deformation of phase-space with the position-position NC parameter $\theta$, and momentum-momentum NC parameter $\eta$, given as
\begin{eqnarray}
	[\hat{\tilde{x}}_1^K,\hat{\tilde{x}}_2^K]=i\theta,\; [\hat{\tilde{p}}_1^K,\hat{\tilde{p}}_2^K]=i\eta,\;
	[\hat{\tilde{x}}_a^K,\hat{\tilde{p}}_b^K]=i\hbar_e\delta_{ab};\; K=A,B.
\end{eqnarray}
The non-symplectic transformation $S$ is a Darboux transformation that connects the NC space and the usual commutative space.
The covariance matrix transforms under $\Upsilon$ as
\begin{equation}\label{sigmatransbopp}
	\Sigma_{AB}\mapsto \tilde{\Sigma}_{AB}=\Upsilon\Sigma_{AB}\Upsilon^T.
\end{equation}
The modified RSUP \cite{bastos2}
\begin{equation}
\tilde{\Sigma}_{AB}+\frac{i}{2}\tilde{\Omega}\ge 0
\end{equation}
can be stated in terms of the spectrum of $2i\tilde{\Omega}^{-1}\tilde{\Sigma}_{AB}$. Since $\tilde{\Omega}^{-1}\tilde{\Sigma}_{AB} = (\Upsilon^{-1})^T \Omega^{-1} \Sigma \Upsilon^T $ and $\Omega^{-1}\Sigma$ are similar, the generalized symplectic spectrum remains unchanged. In particular, for all allowed values of $ m,n,\theta,\eta$, we have
\begin{equation}
	\lambda_{\tilde{\Sigma},min}=b\sqrt{1-R^2}>1,
\end{equation}
which is same as ~\eqref{RSUPspec}. That means the covariance matrix $\Sigma_{AB}$, which satisfies the PPT criteria is transformed into another bonafide covariance matrix $\tilde{\Sigma}_{AB}$ under the Darboux transformation. Now we shall consider two situations- first, we artificially held the transpose operation fixed and secondly, we consider the PT operation that transformed under congruence.
\subsubsection{PT operation remains fixed:}
Let us consider that the NC-deformation changes the canonical structure while  the PPT operation $\Lambda$ held fixed.
To envisage the separability, we perform PT operation ~\eqref{PToperation} on the party $B$. The partial transposition transforms $\tilde{\Sigma}_{AB}$ to $\tilde{\Sigma}'_{AB}=\Lambda\tilde{\Sigma}_{AB}\Lambda^T$. Now we consider only position-position noncommutativity ($\theta\neq 0$ and $\eta=0$). For $\eta=0$, four symplectic invariants for $2i\tilde{\Omega}^{-1}\tilde{\Sigma}'_{AB}$ are explicitly given by
\begin{equation}
	\lambda_{\tilde{\Sigma}'}=\left\{b\sigma_{s1},b\sigma_{s2},b\sigma_{s3},b\sigma_{s4}\right\},\mbox{with}\; \sigma_{sj}=\sqrt{\lambda_{sj}};\; j=1,..,4.
\end{equation}
Explicitly written
\begin{eqnarray}
	\lambda_{s1}= \lambda_{s10} +\frac{1}{2}\sqrt{\lambda_{s11}} +\frac{1}{2}\sqrt{\lambda_{s12} - \frac{\lambda_{s14}}{\sqrt{\lambda_{s11}}}},\\
		\lambda_{s2} = \lambda_{s10} + \frac{1}{2}\sqrt{\lambda_{s11}}-\frac{1}{2}\sqrt{\lambda_{s12}-\frac{\lambda_{s14}}{\sqrt{\lambda_{s11}}}} \\
			\lambda_{s3} = \lambda_{s10} - \frac{1}{2}\sqrt{\lambda_{s11}}+\frac{1}{2}\sqrt{\lambda_{s12}+\frac{\lambda_{s14}}{\sqrt{\lambda_{s11}}}}\\
				\lambda_{s4} = \lambda_{s10} - \frac{1}{2}\sqrt{\lambda_{s11}} - \frac{1}{2}\sqrt{\lambda_{s12}+\frac{\lambda_{s14}}{\sqrt{\lambda_{s11}}}}
\end{eqnarray}
Where
\begin{eqnarray}
	\lambda_{s10} &=& 1+\theta^2+R^2, \\
	\lambda_{s11} &=& 4\theta^2(1+\theta^2+3R^2)+8R^2(1+\sqrt{1+\theta^2(1-R^2)}),\\
			\lambda_{s12} &=& 4\theta^2(1+2\theta^2+7R^2)+8R^2(1-\sqrt{1+\theta^2(1-R^2)}),\\
	\lambda_{s14} &=& -16\theta^2(\theta^2+4R^2)(1+\theta^2+R^2). 
\end{eqnarray}
Accordingly, the minimum symplectic eigenvalue is given by
\begin{equation}\label{lambdasmin}
	\lambda_{s,min}=1+\theta^2+R^2+\theta\sqrt{\theta^2+4R^2}-\sqrt{(1+\theta^2 +R^2+ \theta\sqrt{\theta^2 +4R^2})^2 - (1-R^2)^2}.
\end{equation}
As a consistency check, we examine the undeformed limit of the preceding spectral expressions. Setting $\theta\to 0$, the four positive symplectic eigenvalues reduce to two doubly degenerate values, $b(1+R)$ and $b(1-R)$, in agreement with the canonical PT spectrum obtained in Eq.~\eqref{PTspec}. In particular, in commutative limit, Eq.~\eqref{lambdasmin} yields $\lambda_{s,\min}=(1-R)^2$, so that the minimum positive eigenvalue becomes $b(1-R)=1+R$, in agreement with the canonical limit.
The separability of states requires
\begin{equation}
\sigma_{smin}=	b\sqrt{\lambda_{s,min}}\ge 1.
\end{equation}
\begin{figure}
	\includegraphics[width=8cm]{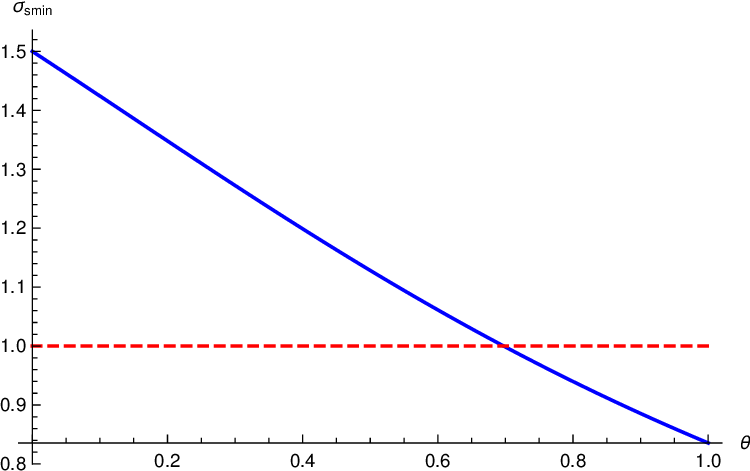}
	\caption{Dependence of the minimum spectral quantity associated with the fixed partial-transpose prescription on the position-position noncommutativity parameter $\theta \in (0,1)$. Here we consider $\eta=0$. We choose the parameter value $R=1/2$. The horizontal line indicates the conventional unit threshold. The crossing illustrates the deformation dependence of the fixed-prescription spectrum.  }
	\label{sigmasmin}
\end{figure}
In the Figure: FIG.\ref{sigmasmin}, we plot the smallest Williamson invariant $\sigma_{smin}$ with respect to the position-position NC-parameter $\theta\in (0,1)$. 
FIG.~\ref{sigmasmin} illustrates the dependence of the minimum spectral quantity obtained using the fixed PT matrix on the position-position noncommutativity parameter $\theta$. For sufficiently large values of $\theta$, this quantity crosses the conventional unit threshold. However, the fixed PT matrix does not generally implement the physical transposition of the deformed operator algebra. Consequently, the observed crossing cannot be interpreted as the generation of physical entanglement. As demonstrated in the following subsection, if we transform PT-operation under congruence accordingly, it preserves the physical PPT classification and the corresponding generalized symplectic spectrum under a local invertible Darboux transformation.
\subsubsection{PT operation changed under congruence:}
The physical PT is an operation on Bob's operator algebra. The matrix representing the operation does depend on the co-ordinates. In particular, under congruence 
\begin{equation}\label{Lambdatrans}
\Lambda\mapsto\tilde{\Lambda}=\Upsilon \Lambda \Upsilon^{-1}=\mathbb{I}_4 \oplus \tilde{\Lambda}_B,\; \mbox{with}	\;\tilde{\Lambda}_B= \Upsilon_B \Lambda_B \Upsilon_B^{-1}.
\end{equation}
Note that, since $\tilde{\Lambda}_B$ is an involution ($\tilde{\Lambda}_B^2 =\mathbb{I}_4 $) and it reverses the NC commutation form as
\begin{equation}
\tilde{\Lambda}_B \tilde{\Omega}_B\tilde{\Lambda}_B^T = - \tilde{\Omega}_B,
\end{equation}
it is a bonafide PT-transformation in NC-space. Using ~\eqref{PTcan},~\eqref{sigmatransbopp} and ~\eqref{Lambdatrans}, we now write the transformation of the covariance matrix in NC-space under PT-transformation as
\begin{equation}
	\tilde{\Sigma}_{AB}' = \tilde{\Lambda}\tilde{\Sigma}_{AB}\tilde{\Lambda}^T = \Upsilon \Sigma_{AB}' \Upsilon^T.
\end{equation}
To be a bonafide covariance matrix after PT, $\tilde{\Sigma}_{AB}'$ has to satisfy the PPT condition
\begin{equation}\label{PPTinNC}
	\tilde{\Sigma}_{AB}' +\frac{i}{2}\tilde{\Omega}\ge 0.
\end{equation}
On the other hand, we can write
\begin{equation}
\tilde{\Sigma}_{AB}' +\frac{i}{2}\tilde{\Omega} = \Upsilon (\Sigma_{AB}'+\frac{i}{2}\Omega)\Upsilon^T ,
\end{equation}
which means the two are connected by an invertible real congruence transformation, thus preserves positive semidefiniteness by virtue of Sylvester's law of inertia \cite{inertia1,inertia2,inertia3}. In other words, the PPT tests in NC space and canonical space are equivalent. Entanglement can not be generated through merely a congruence transformation in phase-space. In particular, Gaussian PPT is covariant under local congruence transformation.\\
In the following section, we demonstrate the situation for pure state in NC-deformation case with an explicit example of two-dimensional anisotropic oscillator.
\section{Toy model: pure state for two dimensional anisotropic oscillator}
We consider a two-dimensional anisotropic harmonic oscillator formulated in NC phase space. To investigate its bipartite Gaussian correlations, we employ the canonical Darboux representation and define the two subsystems through the commuting canonical mode algebras generated by $(\hat{ x}_1, \hat{p}_1)\in \mathcal{H}_1$ and $(\hat{x}_2,\hat{p}_2)\in \mathcal{H}_2$, respectively. The corresponding Hilbert space admits the the decomposition $\mathcal{H}=\mathcal{H}_1\otimes\mathcal{H}_2$. Although the NC position operators do not generally commute, the canonical variables provide a well-defined bipartition for studying the separability of the oscillator ground state. The NC deformation parameters enter the equivalent canonical Hamiltonian through its effective masses, frequencies, and intermode couplings.
Suppose the NC- space  Hamiltonian is given by
\begin{equation}\label{NCShamiltonian}
	\hat{H}_{nc}=\frac{1}{2} \hat{\tilde{X}}^T \mathcal{H}_{nc}\hat{\tilde{X}},\; \mbox{with}\; \hat{\tilde{X}}= (\hat{\tilde{X}}_1,\hat{\tilde{X}}_2, \hat{\tilde{X}}_3, \hat{\tilde{X}}_4)^T= (\hat{\tilde{x}}_1,\hat{\tilde{p}}_1, \hat{\tilde{x}}_2, \hat{\tilde{p}}_2)^T.
\end{equation}
 The anisotropic oscillator in the NC-space is characterized by   the mass $M=(m_1,m_2)$, and the frequency $\tilde{\omega}=(\tilde{\omega}_1,\tilde{\omega}_2)$ through
\begin{eqnarray}
	\mathcal{H}_{nc}= \left(\begin{array}{cc}
		\mathcal{H}_{nc}^{(1)} & 0 \\
		0 & \mathcal{H}_{nc}^{(2)}
	\end{array}\right),\;
	\mbox{with}\;
	\mathcal{H}_{nc}^{(j)}=\mbox{Diag}(m_j\tilde{\omega}_j^2,1/m_j),\; j=1,2.
\end{eqnarray}
We consider the fundamental commutation relations  in NC-space  as 
\begin{equation}\label{NCspacecommutation}
	[	\hat{\tilde{X}}_{\alpha}, \hat{\tilde{X}}_{\beta}]= i \tilde{\Omega}_{\alpha\beta},
\end{equation}
where the deformed symplectic matrix $\tilde{\Omega}$ and the effective Planck constant $\hbar_e$ reads
\begin{eqnarray}
	\tilde{\Omega} = \left(\begin{array}{cc}
		J_2 &  \frac{1}{\hbar_e}\Pi_{\theta\eta}\\
		-\frac{1}{\hbar_e}\Pi_{\theta\eta} &  J_2
	\end{array}\right),\; \mbox{with}\; 
	\Pi_{\theta\eta}= \left(\begin{array}{cc}
		\theta &  0\\
		0 & \eta
	\end{array}\right),\; 
	\hbar_{e}=1+\theta\eta/4.
\end{eqnarray}
Here $\theta$ and $\eta$ are the position position NC-parameter and momentum-momentum NC-parameter, respectively.  $J_2$ is the usual symplectic matrix, which encodes the commutation relations in the usual commutative space 	$\hat{X}=(\hat{X}_1, \hat{X}_2,\hat{X}_3,\hat{X}_4)^T=(\hat{x}_1, \hat{p}_1,\hat{x}_2,\hat{p}_2)^T$ as follows.
\begin{equation}\label{Cspacecommutation}
	[	\hat{X}_{\alpha}, \hat{X}_{\beta}]=i \Omega_{\alpha\beta}, \;
	\mbox{with}\;
	\Omega=\mbox{diag}(J_2,J_2),\; J_2=i\sigma_y.
\end{equation}
The NC-space co-ordinates ($\hat{\tilde{X}}$) are connected to the  commutative space co-ordinates ($\hat{X}$) through the Darboux transformation ($\Upsilon_D$) given by the Bopp's shift
\begin{eqnarray}\label{cncconnection}
	\hat{\tilde{X}}=\Upsilon_D \hat{X},\;\;
	\mbox{with}\;
	\Upsilon_D =\left(\begin{array}{cc}
		\mathbb{I}_2 & -\frac{1}{2}\Pi_{\theta\eta}J_2 \\
		\frac{1}{2}\Pi_{\theta\eta}J_2 & \mathbb{I}_2 
	\end{array}\right).
\end{eqnarray}
The notation $\mathbb{I}_n$ stands for $n\times n$ identity matrix.
Here our concern is to have a valid Darboux transformation. Thus we restrict the parameter choice such that $\theta\eta/4 <1$. In other words, we restrict the determinant of $\Upsilon_D $ is nonzero ($\Delta_{\Upsilon_D}\neq 0$); i.e., $\Upsilon_D \in GL(4, R)$.
$\Omega$ is connected with the deformed symplectic matrix $\tilde{\Omega}$ through 
\begin{equation}\label{JncJcconnection}
	 \tilde{\Omega} =  \Upsilon_D \Omega\Upsilon_D^T .
\end{equation}
Since the quantum mechanical formalism are well established in commutative space, it is customary to convert the NC-space system in the usual commutative space system through ~\eqref{cncconnection} for computational purpose.
Using ~\eqref{cncconnection}, one can see that the NC-space Hamiltonian ~\eqref{NCShamiltonian} is equivalent to the usual commutative space Hamiltonian 
\begin{eqnarray}\label{Hamiltonianmatrix}
	\hat{H}=\frac{1}{2} \hat{X}^T \mathcal{H}\hat{X},\; \mbox{with}\; \mathcal{H}=\Upsilon_D^T \mathcal{H}_{nc}\Upsilon_D.
\end{eqnarray}
Explicitly written
\begin{eqnarray}
	\mathcal{H}=\left(\begin{array}{cc}
		\mathcal{H}_1 & \mathcal{H}_{12}\\
		\mathcal{H}_{12}^T & \mathcal{H}_2
	\end{array}\right),\;
	\mbox{with}\; \mathcal{H}_{12}=\left(\begin{array}{cc}
		0 & -2\nu_2 \\
		2\nu_1 & 0
	\end{array}\right),\; \mathcal{H}_j=\mbox{Diag}(\mu_j\omega_j^2,1/\mu_j); j=1,2
	.
\end{eqnarray}
Explicit forms of the  parameters are given by
\begin{eqnarray}
	\frac{1}{\mu_1} &=& \frac{1}{m_1}+ \frac{1}{4} m_2\tilde{\omega}_2^2\theta^2, \;
	\frac{1}{\mu_2} = \frac{1}{m_2}+ \frac{1}{4} m_1\tilde{\omega}_1^2\theta^2,\\
	\alpha_1&=& \mu_1\omega_1^2=m_1\tilde{\omega}_1^2 + \frac{\eta^2}{4 m_2} ,\;
	\alpha_2  = \mu_2\omega_2^2=m_2\tilde{\omega}_2^2 + \frac{\eta^2}{4 m_1},\\
	\nu_1 & =& \frac{1}{4m_1} (\eta + m_1 m_2\tilde{\omega}_2^2 \theta),\;
	\nu_2 = \frac{1}{4m_2} (\eta + m_1 m_2\tilde{\omega}_1^2 \theta).
\end{eqnarray}
Note that for isotropic oscillator $\mu_1=\mu_2,\; \alpha_1=\alpha_2,\; \nu_1=\nu_2$. For an isotropic oscillator, one can identify the off-diagonal terms constitute the angular momentum operator. \\
One can diagonalize the bilinear Hamiltonian ~\eqref{Hamiltonianmatrix} with a normal coordinate system, keeping the symplectic structure $Sp(4,\mathbb{R})$ intact. First we note that, he symplectic eigenvalues of $\mathcal{H}$ are the ordinary eigenvalues of $
\mathcal{H}_J=\Omega\mathcal{H}$.
Since $	\mathcal{H}_J$ is not symmetric, the left and right eigenvectors of $	\mathcal{H}_J$ are not the same. However, left and right eigenvalues are identical. The characteristic polynomial $	P_{\mathcal{H}_J}(\lambda)$ of $\mathcal{H}_J$ has four distinct purely imaginary roots for the  real parameters $\mu_j,\omega_j,\nu_j$:
\begin{equation}\label{lambdaformexplicit}
	\lambda \in \{\mp i\lambda_j, j=1,2\vert \lambda_1=\sqrt{(\Delta-D)/2}, \lambda_2=\sqrt{(\Delta+D)/2} \},
\end{equation}
where $\Delta=\Delta_{\mathcal{H}_1}+\Delta_{\mathcal{H}_2}+2\Delta_{\mathcal{H}_{12}}$, and the discriminant $D=\sqrt{\Delta^2-4\Delta_{\mathcal{H}_J}}$ is given by
\begin{eqnarray}\label{discriminant}
	D^2= (\omega_1^2-\omega_2^2)^2 + 16\nu_1\nu_2 (\omega_1-\omega_2)^2 + 16 \left(\sqrt{\frac{\mu_1}{\mu_2}}\omega_1\nu_1 + \sqrt{\frac{\mu_2}{\mu_1}}\omega_2\nu_2\right)^2\ge 0.
\end{eqnarray}
Here we have used the notation $\Delta_A=\mbox{Det}(A)$.
Note that $D=0$ only for parameter values $\omega_1=\omega_2,\; \nu_1=\nu_2=0$, which corresponds to the isotropic oscillator in commutative space. We shall consider $D>0$ for our present study.

If $\chi_{lj}$ is the left eigenvector corresponding to the eigenvalue $-i\lambda_j$ of $\mathcal{H}_J$, i.e., $\chi_{lj}\mathcal{H}_J=-i\lambda_j \chi_{lj}$, then the direct computation gives
\begin{eqnarray}
	\chi_{lj}=k_{j}(i \kappa_{j,1}, \kappa_{j,2}, \kappa_{j,3}, i\kappa_{j,4});\; j=1,2.
\end{eqnarray}
Here $k_{j}$ is the normalization constant, and the real parameters $\kappa_{j,\alpha}^{r} $ and $\kappa_{j,\alpha}$ reads
\begin{eqnarray}
	\kappa_{j,1} &=& -2\mu_1\lambda_j (\mu_1\nu_1\omega_1^2 + \mu_2\nu_2\omega_2^2),\;
	\kappa_{j,2} = 2(\mu_2\nu_2\omega_2^2 - 4\mu_1\nu_1^2 \nu_2 + \mu_1\nu_1\lambda_j^2),\\
	\kappa_{j,3} &=& \mu_1(4\mu_1\nu_1^2\omega_1^2 - \mu_2 \omega_1^2 \omega_2^2 + \mu_2\omega_2^2 \lambda_j^2),\;
	\kappa_{j,4} = -\mu_1\lambda_j (\omega_1^2 + 4\nu_1\nu_2 -\lambda_j^2).
\end{eqnarray}
The left eigenvector corresponding to the eigenvalue $i\lambda_j$ is given by $	\chi_{lj}^*$. The right eigenvector $\chi_{rj}$ corresponding to the eigenvalue $-i\lambda_j$   may be obtained through   $\chi_{rj}=-\Sigma_y \chi_{lj}^\dagger$. Normalization condition $\chi_{lj}\chi_{rj}=1$ yields
\begin{equation}\label{normalizationofchi}
	\vert k_{j}\vert= 1/\sqrt{2(\kappa_{j,3}\kappa_{j,4}- \kappa_{j,1} \kappa_{j,2})};\; j=1,2.
\end{equation}
Similarity transformation which diagonalizes $\mathcal{H}_J$, i.e., 
\begin{equation}
	\mathcal{H}_{JD}=\mbox{diag}(-i\lambda_1,i\lambda_1,-i\lambda_2,i\lambda_2)= Q^{-1}\mathcal{H}_J Q,
\end{equation}
is given by the following matrices
\begin{equation}
	Q=(\chi_{r1},\chi_{r1}^*,\chi_{r2},\chi_{r2}^*),\;	Q^{-1}=(\chi_{l1}^T,\chi_{l1}^{*T},\chi_{l2}^T,\chi_{l2}^{*T})^T.
\end{equation}
The diagonal representation of $\mathcal{H}_J$ enables to  define the normal co-ordinates through
\begin{equation}
	\hat{A}=(\hat{a}_1,\hat{a}_1^\dagger,\hat{a}_2,\hat{a}_2^\dagger)^T=Q^{-1}\hat{X}.
\end{equation}
Using the normalization condition ~\eqref{normalizationofchi}, it follows 
\begin{equation}\label{algebraaadagger}
	[\hat{a}_1,\hat{a}_1^\dagger]=	[\hat{a}_2,\hat{a}_2^\dagger]=1.
\end{equation}
On the other hand, the orthogonality condition $\chi_{l1}\chi_{r2}=\chi_{l2}\chi_{r1}=0$ is equivalent to
\begin{equation}\label{algebraa1a2}
	[\hat{a}_1,\hat{a}_2]=0.
\end{equation}
The algebra ~\eqref{algebraaadagger} and ~\eqref{algebraa1a2} confirm that $\hat{a}_j$ and $\hat{a}_j^\dagger$ are annihilation and creation operators.
The ground state of the system thus satisfy the property
\begin{equation}\label{groundstatedefnforanisotropic}
	\hat{a}_1\vert 0,0\rangle =\hat{a}_2\vert 0,0\rangle =0.
\end{equation}
In position representation ($\psi_{0,0}(x_1,x_2)=\langle x_1,x_2\vert 0,0\rangle$), the equation ~\eqref{groundstatedefnforanisotropic} reads
\begin{equation}\label{groundstateforanisotropic}
	(U_xx-i\hbar U_p\partial_{x})\psi_{0,0}(x_1,x_2)=0,
\end{equation}
where
\begin{eqnarray}
	x=\left(\begin{array}{c}
		x_1 \\
		x_2
	\end{array}\right),\; 
	\partial_x=\left(\begin{array}{c}
		\frac{\partial}{\partial x_1} \\
		\frac{\partial}{\partial x_2}
	\end{array}\right),\;
	U_x=\left(\begin{array}{cc}
		i\kappa_{11} & \kappa_{31}\\
		i\kappa_{12} & \kappa_{32}
	\end{array}\right),\;
	U_p=\left(\begin{array}{cc}
		\kappa_{21} & i\kappa_{41}\\
		\kappa_{22} & i\kappa_{42}
	\end{array}\right).
\end{eqnarray}
We take the following ansatz for the solution of ~\eqref{groundstateforanisotropic}.
\begin{equation}\label{ansatzpsi00anisotropic}
	\psi_{0,0}(x_1,x_2)= \mathcal{N}_0 e^{-\frac{1}{2}x^T\Lambda_{ancs}x}.
\end{equation}
$\mathcal{N}_0$ being the normalization condition and
\begin{eqnarray}
	\Lambda_{ancs}=\left(\begin{array}{cc}
		\Lambda_{11} & \Lambda_{12}\\
		\Lambda_{12} & \Lambda_{22}
	\end{array}\right).
\end{eqnarray}
. Using ~\eqref{ansatzpsi00anisotropic} in ~\eqref{groundstateforanisotropic} we get
\begin{equation}
	\Lambda_{ancs}=i U_p^{-1}U_x.
\end{equation}

Explitly written,
\begin{eqnarray}
	\Lambda_{11} &=& \frac{\kappa_{41}\kappa_{12}-\kappa_{42}\kappa_{11}}{\kappa_{21}\kappa_{42}-\kappa_{22}\kappa_{41}} =\Lambda_{11r},\;\;
	\Lambda_{22} = \frac{\kappa_{21}\kappa_{32}-\kappa_{22}\kappa_{31}}{\kappa_{21}\kappa_{42}-\kappa_{22}\kappa_{41}}=\Lambda_{22r}, \label{Lambda11form}\\
	\Lambda_{12} &=& \frac{i(\kappa_{42}\kappa_{31}-\kappa_{41}\kappa_{32})}{\kappa_{21}\kappa_{42}-\kappa_{22}\kappa_{41}}=\frac{i(\kappa_{12}\kappa_{21}-\kappa_{22}\kappa_{11})}{\kappa_{21}\kappa_{42}-\kappa_{22}\kappa_{41}} \label{Lambda12form}=i\Lambda_{12c}.
\end{eqnarray}
In particular, $\Lambda_{11}$ and $\Lambda_{22}$ are real, whereas $\Lambda_{12}$ is purely imaginary. 
\subsection{Noise matrix and separability criterion}
Wigner distribution ($W$) and the density operator $(\hat{\rho})$ are related
through the definition
\begin{equation}
	W(X_c) =\frac{1}{\pi^2}\int d^2 x' \langle x-x'\vert \hat{\rho} \vert x+x'\rangle e^{2ix'.p},
\end{equation}
where $X_c=(X_1,X_2,X_3,X_4)=(x_1,p_1,x_2,p_2)^T$ is the classical co-ordinate  vector. We define 
\begin{eqnarray}
	\Delta \hat{X}_\alpha &=& \hat{X}_\alpha -\langle \hat{X}_\alpha\rangle_\rho,  \mbox{with}\; \langle \hat{X}_\alpha\rangle_\rho = \mbox{Tr}(\hat{X}_\alpha\hat{\rho}); \;\alpha=1,2,3,4.\\
	\Delta X_\alpha &=& X_\alpha -\langle X_\alpha\rangle_W,\; \mbox{with}\; \langle X_\alpha\rangle_W = \int   W(X_c)X_\alpha d^4X_c .
\end{eqnarray}
The four components $\Delta \hat{X}$ satisfy the same commutation relations as $\hat{X}$. Moreover, the phase-space average $\langle X_\alpha\rangle_W$ with respect to the Wigner distribution $W$ is equal to the average $\langle \hat{X}_\alpha\rangle_\rho$ with respect to the density operator $\hat{\rho}$. We define the covariance matrix $\mathcal{V}_c$ through the matrix elements
\begin{equation}
	\mathcal{V}_{\alpha\beta}=\frac{1}{2} \langle \{\Delta \hat{X}_\alpha, \Delta \hat{X}_\beta\}\rangle =\mbox{Tr}(\frac{1}{2}\{\Delta \hat{X}_\alpha, \Delta \hat{X}_\beta\}\hat{\rho})=\int \Delta X_\alpha \Delta X_\beta W(X_c)d^4X_c.
\end{equation}
Since $\mathcal{V}_c$ is symmetric, we can  write it in the block form
\begin{eqnarray}\label{cspacecovariancematrix}
	\mathcal{V}_c=\left(\begin{array}{cc}
		V_{11} & V_{12}\\
		V_{12}^T & V_{22}
	\end{array}\right).
\end{eqnarray}
Using the connection ~\eqref{cncconnection} between the commutative space and NC-space co-ordinates, one can see that the NC-space covariance matrix $\mathcal{V}_{nc}$ is given by
\begin{equation}\label{VncVcconnection}
	\tilde{\mathcal{V}}_{nc}= \Upsilon_D  \mathcal{V}_c\Upsilon_D^T .
\end{equation}
The fundamental commutation relations ~\eqref{Cspacecommutation} of commutative space implies that a bonafide covariance matrix must satisfy the Robertson-Schr\"{o}dinger uncertainty principle (RSUP)
\begin{equation}\label{RSUPc}
	\mathcal{V}_c+\frac{i}{2} \Omega \ge 0 .
\end{equation}
In NC-space, the equivalent statement for the RSUP reads
\begin{equation}\label{RSUPnc}
	\tilde{\mathcal{V}}_{nc} + \frac{i}{2}  \tilde{\Omega} \ge 0.
\end{equation}
A generic local transformation $S_1 \bigoplus S_2$, acts on $\mathcal{V}_c$ as
\begin{equation}
	V_{jj}\to S_j V_{jj} S_j^T ,\; 	V_{12}\to S_1 V_{12}S_2^T;\; \mbox{with}\; S_j \in Sp(2,\mathbb{R}),\;  j=1,2.
\end{equation}
One can identify that following four quantities are local  invariant with respect to transformation belonging to the $Sp(2,\mathbb{R})\bigotimes Sp(2,\mathbb{R}) \subset Sp(4,\mathbb{R})$.
\begin{eqnarray}
	\Delta_j=Det(V_{jj}),	\Delta_{12}=Det(V_{12}),  \Delta_{\mathcal{V}_c} = Det(\mathcal{V}_c),
	\tau_{v}=\mbox{Tr}(V_{11} J_2 V_{12} J_2 V_{22} J_2 V_{12}^T J_2).
\end{eqnarray}
Using Williamson's theorem \cite{Williamson1,Williamson2}, one can show that the RSUP ~\eqref{RSUPc} can be rewritten as $Sp(2,\mathbb{R})\bigotimes Sp(2,\mathbb{R})$ invariant statement
\begin{equation}\label{covariancersupc}
	\Delta_1 \Delta_2 + (1/4- \Delta_{12})^2 -\tau_{v}\ge ( \Delta_1+\Delta_2)/4.
\end{equation}
Under mirror reflection (Peres-Horodecki partial transpose) $\Delta_1$,  $\Delta_2$ and $\tau_v$ remain invariant; whereas $\Delta_{12}$ flips sign. Therefore, the requirement that the covariance matrix of a separable state has to obey the following necessary condition.
\begin{equation}\label{separabilityc}
	Ps=	\Delta_1 \Delta_2 + (1/4-\vert \Delta_{12}\vert)^2 - \tau_{v}- ( \Delta_1+\Delta_2)/4 \ge 0,
\end{equation}
which turns out to be sufficient for all bipartite-Gaussian state.\\
Now, for a generic ground state ~\eqref{groundstatedefnforanisotropic}, which has the form in position representation as
$\psi(x_1,x_2)=\mathcal{N}_0e^{-\frac{1}{2}x^T\Lambda_{ancs} x}$, corresponds to 
the Wigner distribution
\begin{equation}\label{WX}
	W(X)= \frac{1}{\pi^2} \exp\{ - x^T (\Lambda_r + \Lambda_c \Lambda_r^{-1}\Lambda_c^T)x -  p^T\Lambda_r^{-1}p  - (x^T \Lambda_c \Lambda_r^{-1} p + p^T \Lambda_r^{-1}\Lambda_c^T x) \}.
\end{equation}
Here the following notation for the matrix $\Lambda_{ancs}$ and its elements $\Lambda_{jk}$ have been used.
\begin{eqnarray}
	\Lambda_{ancs}=[\Lambda_{jk}]_{j,k=1}^{2}= \Lambda_r +i\Lambda_c= [\Lambda_{jkr}]_{j,k=1}^{2}+i[\Lambda_{jkc}]_{j,k=1}^{2},
\end{eqnarray}
where $\Lambda_{jkr}=\Re(\Lambda_{jk}),\Lambda_{jkc}=\Im(\Lambda_{jk})$, and $x=(x_1,x_2)^T,\; p=(p_1,p_2)^T$.
The expectation values through the Wigner distribution $W(X)$, provide the covariance matrix $\mathcal{V}_{c}$ as
\begin{eqnarray}\label{vc}
	\mathcal{V}_{c}=  \frac{1}{2}\left(\begin{array}{cc}
		\sigma_{11} & \sigma_{12}\\
		\sigma_{12}^T & \sigma_{22}
	\end{array}\right),
\end{eqnarray}
with
\begin{eqnarray}\label{formofsigmaij}
	\sigma_{11}=\left(\begin{array}{cc}
		\frac{1}{\Lambda_{11r}} & 0\\
		0 & \frac{\Delta_\Lambda}{\Lambda_{22r}}
	\end{array}\right),\;
	\sigma_{22}=\left(\begin{array}{cc}
		\frac{1}{\Lambda_{22r}} & 0\\
		0 & \frac{\Delta_\Lambda}{\Lambda_{11r}}
	\end{array}\right),\;
	\sigma_{12}=\left(\begin{array}{cc}
		0& -	\frac{\Lambda_{12c}}{\Lambda_{11r}} \\
		-\frac{\Lambda_{12c}}{\Lambda_{22r}} &0
	\end{array}\right),
\end{eqnarray}
where $\Delta_\Lambda= \Lambda_{11r}\Lambda_{22r}+\Lambda_{12c}^2$. Note that $\mathcal{V}_c$ satisfies the pure state condition
\begin{equation}
\mathcal{V}_c \Omega\mathcal{V}_c=\frac{1}{4}\Omega \implies \mbox{Det}(\mathcal{V}_c)=1/16,
\end{equation}
as expected.
Using  ~\eqref{formofsigmaij} in the generalized Peres-Horodecki separability criterion ~\eqref{separabilityc} we get the following constraint on the parameters.
\begin{equation}
	-\Lambda_{11r} \Lambda_{22r} \Lambda_{12c}^2 \ge \Lambda_{12c}^2 \Lambda_{11r}\Lambda_{22r}\implies
	\Lambda_{11r}\Lambda_{22r}\Lambda_{12c}=0.
\end{equation}
Since $\Lambda_{11r}$ and $\Lambda_{22r}$ are nonzero, the separability of states implies $\Lambda_{12c}=0$, i.e.,
\begin{equation}\label{constrainteqn}
	(\lambda_2-\lambda_1)(\mu_2\nu_2\omega_2^2 -4\mu_1\nu_1^2\nu_2-\mu_1\nu_1\lambda_1\lambda_2)=0.
\end{equation}
However, according to ~\eqref{discriminant} and ~\eqref{lambdaformexplicit}, we consider $\lambda_1\neq \lambda_2$. Therefore, ~\eqref{constrainteqn} holds for
\begin{equation}
	(\mu_1^2\omega_1^2\nu_1^2-\mu_2^2\omega_2^2\nu_2^2)(\omega_2^2-4\mu_1\nu_1^2/\mu_2)=0.
\end{equation}
From the physical viewpoint, we choose $\theta,\eta \le 1$. In other words, $\theta\eta \neq 4$, which means $\omega_2^2 \neq 4\mu_1\nu_1^2/\mu_2$. Therefore, the only possibility for the separable states is satisfied by the constraint $\mu_1\nu_1\omega_1 =\mu_2\nu_2\omega_2$, which is  trivially satisfied for commutative space ($\theta,\eta\to 0$), and as well as for isotropic oscillator ($\tilde{\omega}_1=\tilde{\omega}_2$) in noncommutative (NC) space. Therefore, the entanglement between the coordinate degrees of freedom is not the sole property of the noncommutativity of space. It depends on both the NC-parameters ($\theta,\eta$) and the anisotropy of oscillator frequency ($\tilde{\omega}_1\neq\tilde{\omega}_2$). If we recall that the off-diagonal term for the isotropic oscillator in NC-deformed space under consideration is merely an angular momentum operator, which commutes with the rest part of Hamiltonian in the equivalent commutative space. Thus, the ground state (Gaussian) is just product of two ordinary oscillators, results into the separable states. However, for anisotropic oscillator, the deformed angular momentum operator does not commute with the rest portion of the Hamiltonian, which results into the entangled state. The canonical representation of the NC oscillator contains deformation-dependent intermode couplings. Consequently, variations of the NC parameters generally modify the effective Hamiltonian and may change the correlations and entanglement of its Gaussian ground state. These changes arise from the parameter dependence of the Hamiltonian and should not be interpreted as entanglement generation by a passive Darboux transformation. In particular, a nonzero intermode coupling does not necessarily imply an entangled ground state. The present analysis only concerns entanglement relative to the bipartition defined by the canonical oscillator modes.
\subsection{Geometry induced by NC deformation for anisotropic oscillator pure state}
The covariance matrix \eqref{vc}, corresponding to the bipartite oscillator state ~\eqref{ansatzpsi00anisotropic}, can be expressed in the normal form
\begin{equation}
	\Sigma_{nf}= \left(\begin{array}{cc}
		a\mathbb{I}_2 & c\sigma_z \\
		c\sigma_z & a \mathbb{I}_2
	\end{array}\right),\;
	\mbox{with}\; a=\frac{1}{2}\sqrt{1+\frac{\Lambda_{12c}^2}{\Lambda_{11r}\Lambda_{22r}}},\; c= \frac{ \Lambda_{12c}}{2\sqrt{\Lambda_{11r}\Lambda_{22r}}}.
\end{equation} 
If we parametrize $(a,c):=(\vartheta_0,\vartheta_1)$, then ~\eqref{gmunuV} provides the  components of metric tensor ($g_{\mu\nu},\; \mu,\nu =0,1$), which reads
\begin{eqnarray}
	g_{00}=g_{11}=\frac{2(a^2+c^2)}{(a^2-c^2)^2},\; 
	g_{01}=g_{10}=-\frac{4ac}{(a^2-c^2)^2}.
\end{eqnarray}
Note that $a$ and $c$ satisfy the pure state condition $a^2-c^2=1/4$. Thus, the oscillator ground states occupy only the constrained submanifold $a^2-c^2=1/4$.
The Fisher information matrix $g(a,c)$ has the eigenvalues $\lambda_{g\pm}=\frac{2}{(a\pm c)^2}$, and $g$ can be diagonalized (locally) with the similarity transformation
\begin{eqnarray}
	g_d=\mbox{diag}(\lambda_{g+},\lambda_{g-})=Q_gg Q_g^T,\;\mbox{with}\; 
	Q_g=\frac{1}{\sqrt{2}}\left(\begin{array}{cc}
		1 & -1 \\
		1 & 1
	\end{array}\right).
\end{eqnarray}
Accordingly, in this diagonal representation of the FR-matrix the co-ordinate vector $(a,c)^T$ is transformed into (locally)
\begin{equation}
	(a,c)\to (u_0,v_0) := (u_0,v_0)^T=Q_g(a,c)^T,
\end{equation}
which reduces the metric to
\begin{equation}
	ds_{FR}^2=\frac{du_0^2}{u_0^2}+\frac{dv_0^2}{v_0^2}.
\end{equation}
Introducing $u_0=e^u$ and $v_0=e^v$, the distance function can be reduced to the metric
\begin{equation}\label{FRpure}
	ds_{FR}^2= du^2+dv^2.
\end{equation}
Thus, the two-dimensional symmetric normal-form covariance manifold is intrinsically flat (Ricci scalar $\mathcal{R}=0$).  Due to the pure state constraint, we can reparametrize $a$ and $c$ as
\begin{equation}
	a=1/2 \cosh(2r), \; c=1/2 \sinh(2r),
\end{equation}
and write the distance function ~\eqref{FRpure} as
\begin{equation}
	ds_{FR,pure}^2 =8dr^2 \implies \mathcal{R}=0.
\end{equation}
Note that, the normal form $\Sigma_{nf}$ is connected with $\mathcal{V}_c$ through local canonical transformation  $S_{loc}=S_1 \bigoplus S_2$, which preserves the canonical bipartition and Gaussian entanglement classification. However, $S_{loc}$, generally depends on covariance parameters. Therefore, the geometry for the original oscillator state might be different. 

Using the parameter  $\vartheta = (\vartheta_1, \vartheta_2, \vartheta_3) = (\Lambda_{11r}, \Lambda_{22r}, \Lambda_{12c})$ in ~\eqref{gmunuV}, the covariance matrix \eqref{vc}, provide the FR-information metric
\begin{equation}
	g(\vartheta)=\mbox{diag}(\vartheta_1^{-2}, \vartheta_2^{-2}, 2\vartheta_1^{-1} \vartheta_2^{-1}).
\end{equation}
Therefore, the FR line element is
\begin{equation}\label{lineel}
	ds_{FR}^2 = \vartheta_1^{-2}d\vartheta_1^2 + \vartheta_2^{-2}d\vartheta_2^2 + 2\vartheta_1^{-1} \vartheta_2^{-1}d\vartheta_3^2.
\end{equation}
If we introduce the co-ordinates $(w,z,c)$ as
\begin{equation}\label{wzc}
	w=\frac{u+v}{\sqrt{2}},\; z=\frac{u-v}{\sqrt{2}},\; c=\vartheta_3,\; \mbox{with}\; u=\ln\vartheta_1,\; v=\ln\vartheta_2,
\end{equation}
then ~\eqref{lineel} can be rewritten as
\begin{equation}\label{FRosc}
	ds_{FR}^2=dw^2+dz^2+2e^{-\sqrt{2}w}dc^2,
\end{equation}
which corresponds to a three-dimensional Riemannian manifold parametrized by the independent Gaussian parameters $\Lambda_{11r}$, $\Lambda_{22r}$, and $\Lambda_{12c}$. Unlike the symmetric normal-form covariance family, this manifold retains the local covariance-width degrees of freedom together with the intermode correlation parameter. The nonvanishing Christoffel symbols for ~\eqref{FRosc} reads
\begin{equation}
\Gamma^w_{cc}=\sqrt{2}e^{-\sqrt{2}w},\; \Gamma^c_{wc}=\Gamma^c_{cw}=-\frac{1}{\sqrt{2}}.
\end{equation}
The corresponding nonzero components of the Ricci tensor are
\begin{equation}
R_{ww}=-\frac{1}{2},\; R_{cc}=-e^{-\sqrt{2}w}.
\end{equation}
 Consequently, the Ricci scalar is $\mathcal R_{osc}=-1$. Thus, the original three-parameter covariance family possesses constant negative scalar curvature.  The negative scalar curvature arises from the coupling between the collective covariance-width parameter $w$ and the intermode correlation parameter $c$. For the oscillator ground-state family, the relation
$\frac{\Lambda_{11r}}{\Lambda_{22r}}=\frac{m_1\tilde{\omega}_1}{m_2\tilde{\omega}_2}=\mbox{constant}$,
 restricts the three-dimensional covariance manifold to a two-dimensional hyperbolic submanifold with induced FR metric
 \begin{equation}
 	ds_{FR,red}^2=dw^2 + 2e^{-\sqrt{2}w}dc^2.
 \end{equation}
Consequently, the Gaussian curvature of the nondegenerate ground-state manifold is $K=-1/2$ (equivalently, the scalar curvature is $\mathcal{R}=-1$), independent of the NC deformation parameters $\theta$  and $\eta$.
 \\
It is important to distinguish this result from the vanishing intrinsic curvature obtained for the symmetric normal-form covariance family. Although the original covariance matrix and its normal form are related by a local symplectic transformation, the transformation generally depends on the covariance parameters. Consequently, it does not necessarily preserve the FR-metric induced on the respective parameterized families. The normal-form description eliminates the local squeezing degrees of freedom and, after imposing the pure-state constraint, reduces to a one-dimensional family characterized by the two-mode squeezing parameter.
The constant negative curvature of the original covariance manifold does not, by itself, indicate quantum entanglement. For the pure oscillator ground state, entanglement is determined by the  correlation parameter $\Lambda_{12c}/\sqrt{\Lambda_{11r}\Lambda_{22r}}$, whereas the scalar curvature remains equal to $\mathcal{R}_{osc}=-1$ throughout the three-dimensional parameter domain, including the separable subset. Therefore, the FR curvature characterizes the statistical geometry of the covariance family rather than serving as a direct entanglement measure.
\section{Conclusions}
In this work, we have investigated the relationship between Fisher-Rao (FR) information geometry and Gaussian quantum correlations under nonsymplectic phase-space transformations, particularly noncommutative (NC) deformation implemented through the Bopp shift. Using the established affine invariance of the FR-distance under invertible congruence transformations, we have examined the distinction between statistical distinguishability and the physical separability of Gaussian states. Our analysis demonstrates that, although FR-distances remain invariant under such transformations, the physical interpretation of Gaussian entanglement additionally requires consistent treatment of the commutation structure, partial transposition, and subsystem decomposition.
\\
For a bipartite Gaussian system, we have shown that retaining a fixed partial-transpose (PT) matrix in the deformed phase space produces deformation-dependent spectral shifts that may cross the conventional positive partial transpose (PPT) threshold. However, these crossings do not represent physical entanglement generation. When the PT operation is consistently transported together with the covariance and commutation matrices, the physical PPT spectrum and classification remain invariant under local Darboux transformations. Thus, a passive local nonsymplectic transformation cannot, by itself, generate entanglement.
\\
We have further investigated a two-dimensional anisotropic oscillator in NC phase space.  The associated FR geometry reveals that the original three-parameter covariance manifold possesses constant negative scalar curvature $\mathcal{R}=-1$, while the nondegenerate two-parameter ground-state manifold has Gaussian curvature $K=-1/2$, independent of the NC deformation parameters. In contrast, the pure symmetric normal-form covariance family is intrinsically one-dimensional and flat. These geometric differences reflect the parameter dependence of the local symplectic transformations used to obtain the normal form.
\\
Our results establish that FR geometry and Gaussian entanglement characterize distinct aspects of the covariance-state description. In particular, negative information-geometric curvature does not necessarily imply quantum entanglement, since it persists even for separable oscillator ground states. The present analysis therefore clarifies the distinction between passive phase-space transformations and deformation-dependent changes in physical quantum states, providing a consistent framework for examining Gaussian correlations and information geometry in NC systems.
\section{Data Availability Statement}
The manuscript has no associated data.
\section{Conflict of interests}
All the authors declare that there is no conflict of interest from funding agency or any other means whatsoever.  
\section{Acknowledgement}
P. Patra is grateful to ANRF (Formerly SERB), Govt. of India, for financial support through project grant No. EEQ/2023/000784. The authors are grateful to anonymous referees for their valuable suggestions, which were helpful in making this article in its present form.

\end{document}